\documentclass[final,preprint,3p,square,times,twocolumn,review]{elsarticle}

\usepackage{lineno,hyperref}
\modulolinenumbers[5]
\usepackage[german,american,english]{babel}
\usepackage{graphicx}
\usepackage{graphics}
\usepackage{dcolumn}
\usepackage{bm}
\usepackage{amssymb}
\usepackage{amsmath}
\usepackage{amsfonts}
\usepackage{epsfig}
\usepackage{mathbbol}
\usepackage{dcolumn}
\usepackage{subfigure}
\biboptions{sort&compress}
\hypersetup{
colorlinks,
citecolor=blue,
filecolor=blue,
linkcolor=blue,
urlcolor=blue
}

\newcommand {\pd} [2] {\frac{\partial #1}{\partial #2}}
\newcommand{\dps}{\displaystyle}

\newcommand{\om}{\iffalse}
\newcommand{\ba}{\arraycolsep 0.3ex \begin{array}{rl}}
\newcommand{\ea}{\end{array}}

\journal{Journal of Physics and Chemistry of Solids}

\begin{document}

\begin{frontmatter}

\title{Coulomb Drag in Topological Materials}

\author{Hong Liu}
\author{Dimitrie Culcer}
\address{School of Physics and Australian Research Council Centre of Excellence in Low-Energy Electronics Technologies, UNSW Node, The University of New South Wales, Sydney 2052, Australia}




\begin{abstract}
Dirac fermions are at the forefront of modern condensed matter physics research. They are known to occur in materials as diverse as graphene, topological insulators, and transition metal dichalcogenides, while closely related Weyl fermions have been discovered in other materials. They have been predicted to lend themselves to a variety of technological applications, while the recent prediction and discovery of the quantized anomalous Hall effect of massive Dirac fermions is regarded as a potential gateway towards low-energy electronics. Some materials hosting Dirac fermions are natural platforms for interlayer coherence effects such as Coulomb drag and exciton condensation. The top and bottom surfaces of a thin topological insulator film provide such a prototype system. Here we describe recent insights into Coulomb drag between two layers of Dirac fermions relying primarily on topological insulator films as a minimal model. We consider both non-magnetic topological insulators, hosting massless Dirac fermions, and magnetic topological insulators, in which the fermions are massive. We discuss in general terms the dynamics of the thin-film spin density matrix, outlining numerical results and approximate analytical expressions where appropriate for the drag resistivity $\rho_\text{D}$ at low temperatures and low electron densities. In magnetic topological insulators with out-of-plane magnetizations in both the active and passive layers we analyze the role of the anomalous Hall effect in Coulomb drag. Whereas the transverse response of the active layer is dominated by a topological term stemming from the Berry curvature, we show that neither the topological mechanism nor disorder renormalizations associated with it contribute to Coulomb drag. Nevertheless, the longitudinal drag force in the passive layer does give rise to a transverse drag current that is independent of the active-layer magnetization. It depends non-monotonically on the passive-layer magnetization, exhibiting a peak that becomes more pronounced at low densities. All of these observations can be verified in the laboratory. We compare results for topological insulators with results for graphene, identifying qualitative and quantitative differences, and discuss generalisations to multi-valley systems, ultra-thin films and electron-hole layers.
\end{abstract}

 
\end{frontmatter}


\section{Introduction}

The energetic study of Dirac fermions in condensed matter physics is currently in full bloom and is driven primarily by the realization of their considerable potential for spin electronics, thermoelectricity, magnetoelectronics and topological quantum computing \citep{Das_Sarma_RMP_TQC}. Exotic physical phenomena associated with Dirac fermions have been identified in materials ranging from graphene \citep{KaneMele_QSHE_PRL05} to topological insulators, \citep{Hasan_TI_RMP10} transition metal dichalcogenides \citep{Dixiao_MoS2_2012} and Weyl and Dirac semimetals. \citep{Burkov_Weyl_TI, Fuhrer_Na3Bi_2016, Fuhrer_NL_Dirac_semi} The linear dispersion of massless Dirac fermions has aroused considerable interest experimentally  \citep{Wang_Bi2Te3_Ctrl_AM11, Benjamin_nature_2011,He_Bi2Te3_Film_WAL_ImpEff_PRL11, Kim_sur_2012, Jinsong-Zhang-Yayu-Wang-2013-science, Lucas_nl_tran_2014, Jianshi_nl_spin_polarized_2014, Kozlov_prl_2014, Fuhrer_nature_2013, Cacho_prl_2015, ZhangJinsong_prb_2015, Hellerstedt_APL_2014,Kastl_surface_tran_nature_2015, E_control_SOT_TI} and theoretically \citep{Hwang_Gfn_Screening_PRB07,JungMacDonald_graphene_PRB2011,Durst_2015,Haizhou_WL_2014,FP_Bi2X_3_NewJ, Shi_Rappe_nl_2016, Culcer_TI_AHE_PRB11,Adam_2D_Tran_prb_2012,Yoshida-PRB-2012,LiQiuzi_2013_prb,Weizhe_TITF_2014_prb,Hai-Zhou_Conductivity_2014_prl,Das_prb_2015}. 
Among Dirac fermion systems, three-dimensional topological insulators (TIs) stand out as a novel class of bulk insulating materials exhibiting conducting surface states with a chiral spin texture\citep{KaneMele_QSHE_PRL05,Hasan_TI_RMP10,Qi_TI_RMP_10,Moore_TRI_TI_Invariants_PRB07,Ando-TI, Culcer_TI_PhysE12, Tkachov_TI_Review_PSS13,Yong-qing_FOP_2012, Moore-nature}, with the surface carriers being massless Dirac fermions. Magnetic topological insulators, hosting massive Dirac fermions, have also been successfully manufactured \citep{Hor_DopedTI_FM_PRB10,Jinsong-Zhang-Yayu-Wang-2013-science,Collins-McIntyre_Cr_Bi2Se3_2014}. Time-reversal symmetry breaking gives Dirac fermions a finite mass and results in a non-trivial Berry curvature \cite{Dixiao_MoS2_2012,LuZhao_TITF_transport_PRL2013, Valley-Hall_graphene_prl_2007, Valley_prb_Niu, Bi-graphene_massive_prl}, which is at the root of the anomalous \citep{Nagaosa-AHE-2010, Culcer_TI_AHE_PRB11} and quantum anomalous Hall effects \citep{Yu_TI_QuantAHE_Science10} detected in these materials \citep{JiangQiao_TIF_QAHE_PRB2012, Chang_TIF_ferromagnetism_AHE_AM2013, Chang_QAHE_exper_Science2013}. In magnetic TI slabs a dissipationless quantized anomalous Hall effect has been discovered  \citep{Murakami_SHI_PRL04, QSH-HgTe_2007, Chang_QAHE_exper_Science2013, Precise-QAHE-CZChang}, which has already been harnessed successfully \citep{D.Reilly_USYD}, stimulating an intense search for device applications. In a related context hybrid TI/superconductor junctions have been fabricated \citep{Sochnikov_TISC_nano_2013,Jin-Feng_TISC_2014}, in which topological superconductivity and Majorana fermions may be present \citep{SDS_TQC_RMP08,Qiang-Hua_TISC_2014_SciRep}.

\begin{figure}
\begin{center}
\includegraphics[width=1\columnwidth]{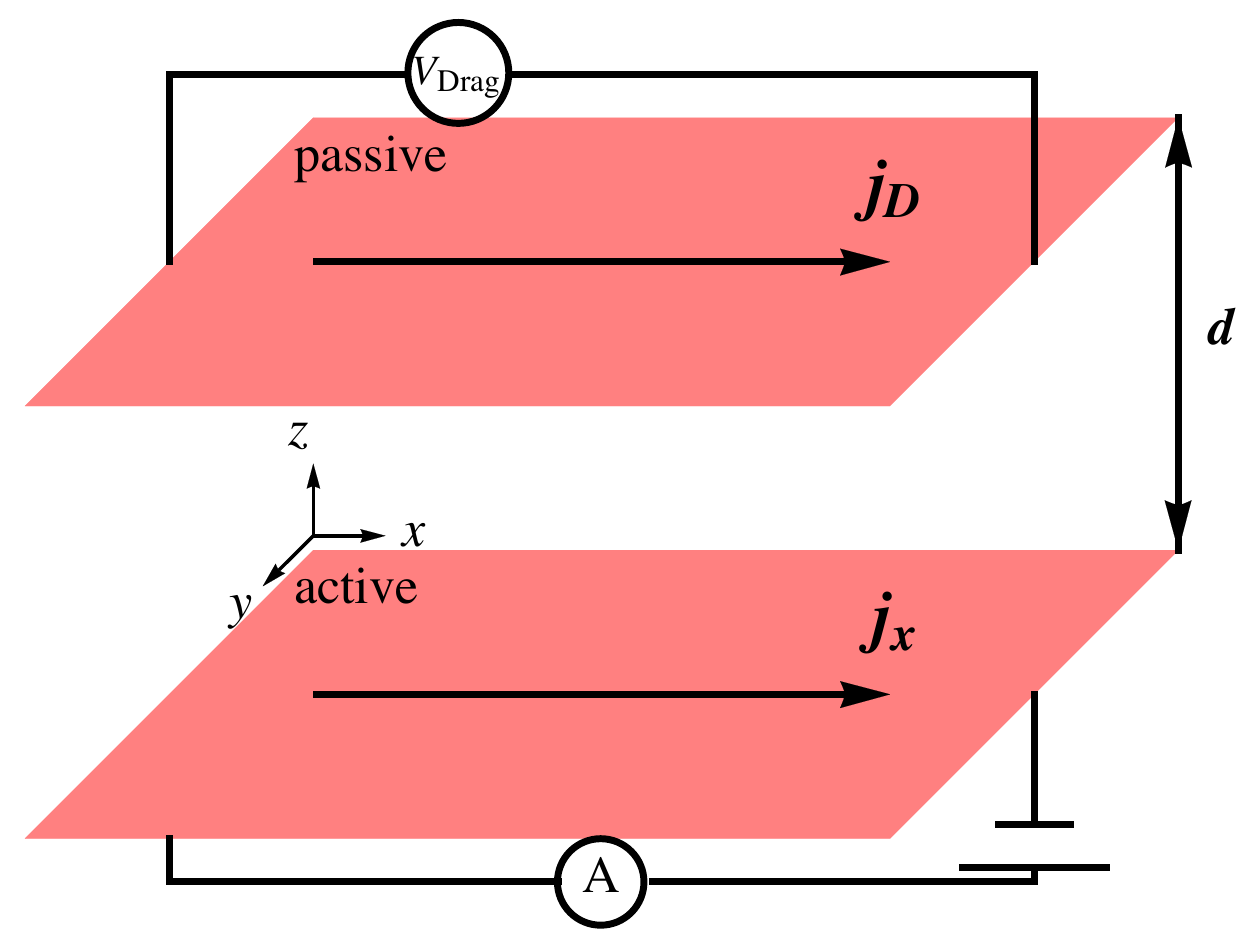}\\
\caption{\label{TI_Drag} An experimental Coulomb drag setup in a non-magnetic topological insulator. The longitudinal current ${\bm j}_x$ in the active layer \textit{drags} a current ${\bm j}_D$ in the passive layer.}
\end{center}
\end{figure}

Three-dimensional topological insulator slabs are inherently two-layer systems naturally exhibiting interlayer coherence effects such as Coulomb drag \cite{1977_drag, Zheng-2DES-1993, Drag_riview}, in which the charge current in one layer \textit{drags} a charge current in the adjacent layer through the interlayer electron-electron interactions, as shown in Fig.~\ref{TI_Drag}. Physically, Coulomb drag is caused by the transfer of momentum between electrons in different layers due to the interlayer electron-electron scattering. It constitutes an experimental probe of electron-electron interactions \citep{1977_drag, 1995_Hall_drag,1999_Hall_drag} and has recently attracted considerable attention in massless Dirac fermion systems such as graphene \citep{Tse_SO_Drag_PRB07, Katsnelson2011, Hwang2011, Tse2007, Amorim_drag, M.Carrega2012, Kim2011, Narozhny_drag_2012, Kim2012, Gorbachevi_nature_2012, Sch_drag_prl_2013, Gamucci_nature_2014, Hall_drag_Graphene,Song_Halldrag_2013}. A significant motivation for the intensive experimental effort dedicated to Coulomb drag in Dirac fermion systems is the search for exciton condensation in electron-hole bilayers \citep{Drag-dot, 1D-1D_science_Luttinger,1D-1D_prb_Dmitriev, edge_drag_prb, ee_drag_93_prb, Nandi_eh_nat, E-H_exciton_2016, Disorder_drag, Drag_semi_1996,Semi_drag_prl, 1995_Hall_drag,drag_metal_prb, Tse2007, Amorim_drag, Kim2011,Tse_SO_Drag_PRB07, Katsnelson2011, Hwang2011, M.Carrega2012, Narozhny_drag_2012, Kim2012, Gorbachevi_nature_2012, Sch_drag_prl_2013, Gamucci_nature_2014, Hall_drag_Graphene, Song_Halldrag_2013, Scharf_Alex_drag_prb_2012, Drag_graphnen_nl, Drag_TI_2013_Efimkin, Hong_drag_2015}. At the same time Coulomb drag of massive Dirac fermions is associated with an important question: if topological terms are present in the drag current they could be exploited in longitudinal transport, potentially enabling a topological transistor. 

In this article we discuss Coulomb drag of both massless and massive Dirac fermions. We take as our prototype model the surface states of 3D topological insulators and we consider two cases: (i) massless Dirac fermions, in which neither layer is magnetized; (ii) massive Dirac fermions, in which an out of plane magnetization exists in each layer, but no external magnetic field. The physics discussed is therefore that of 2D Dirac fermions: the effects are driven by the electrons in the surface conduction band(s) of a topological insulator. Case (i) provides a useful contrast to graphene. Unlike graphene, in TIs the spin and orbital degrees of freedom are coupled by the strong spin-orbit interaction, TIs have an odd number of valleys on a single surface, and the relative permittivity is different, while in known band TIs screening is qualitatively and quantitatively different, since it does not involve the interplay of the layer and valley degrees of freedom. All these features impact the drag current. We introduce a density matrix method to calculate the Coulomb drag current in topological insulator films, which fully takes into account the spin degree of freedom and interband coherence. The drag resistivity in case (i) takes the analytical form
 \begin{equation}\label{rhoD}
 \rho_\text{D}=-\frac{\hbar}{e^2}\frac{\zeta(3)}{16\pi}\frac{(k_\text{B}T)^2}{A^2r^2_sn^{\frac{3}{2}}_\text{a}
 n^{\frac{3}{2}}_\text{p}d^4},
\end{equation}
where $k_\text{B}$ is the Boltzmann constant, $A$ is the TI spin-orbit constant,  $r_s$ is the Wigner-Seitz radius (effective fine structure constant) which represents the ratio of the electrons' average Coulomb potential and kinetic energies, $d$ is the layer separation and $n_{\text{a},\text{p}}$ are the electron densities in the active and passive layers, respectively. For a single-valley system $r_s = e^2/(2\pi\epsilon_0\epsilon_r A)$, with $\epsilon_r$ the relative permittivity. Whereas the full numerical results are more complicated, the same trends are present as in the analytical formula.

\begin{figure}
\begin{center}
\includegraphics[width=1\columnwidth]{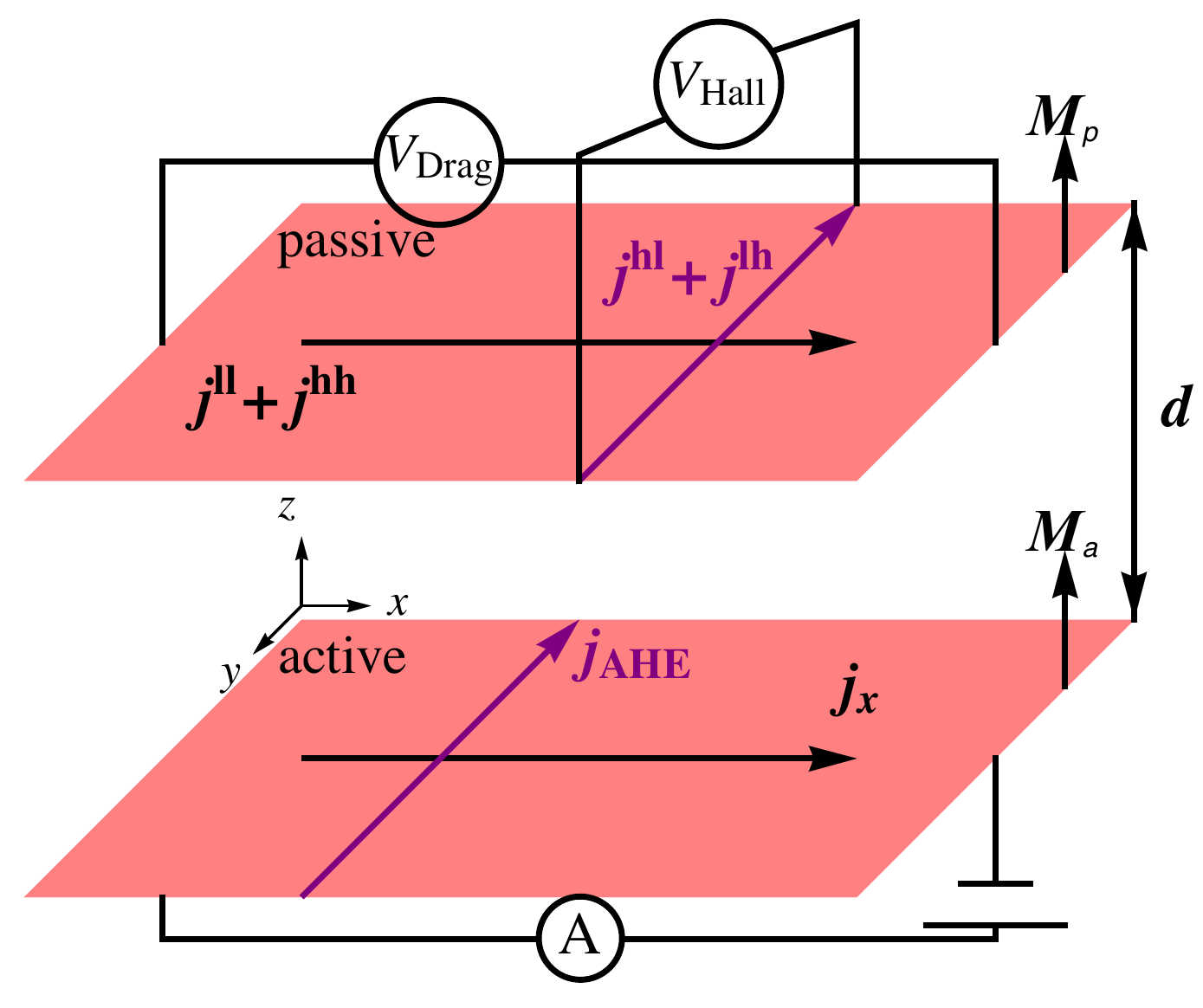}\\
\caption{\label{4-currents} Contributions to the drag current in magnetic TIs. The electric field $\parallel\hat{\bm x}$ gives rise to a longitudinal (${\bm j}_x$) and an anomalous Hall (${\bm j}_{\text{AHE}}$) current in the active layer. In the passive layer there are four contributions to the drag currents: ${\bm j}^{\text{ll}}$ is the longitudinal current dragged by ${\bm j}_x$; ${\bm j}^{\text{lh}}$ is the transverse current dragged by ${\bm j}_{\text{AHE}}$; ${\bm j}^{\text{hl}}$ is the anomalous Hall current generated by ${\bm j}^{\text{ll}}$; while ${\bm j}^{\text{hh}}$ is the anomalous Hall current generated by ${\bm j}^{\text{lh}}$, and it flows longitudinally. ${\bm M}_\text{a}$ and ${\bm M}_\text{p}$ are the magnetizations of the active and passive layers, respectively. $d$ is the layer separation.}
\end{center}
\end{figure}

Case (ii) is significantly more complicated. The drag current in the passive layer has a longitudinal component and a Hall component, which can be measured separately. The Hall drag current in principle has contributions from (i) the longitudinal current in the active layer via a \textit{transverse drag force}, (ii) the Hall current in the active layer, which may be termed \textit{direct Hall drag} (Fig.~\ref{4-currents}). Contribution (i) represents straightforward longitudinal charge transport in the presence of impurities and other scattering mechanisms. Contribution (ii) is topological: the anomalous Hall effect in Dirac fermion systems is predominantly driven by a topological term originating in the Berry curvature associated with the band structure. The Hall drag current stemming from the \textit{transverse drag force} of (i) in effect comes from the conventional longitudinal drag current.

Interestingly, the (topological) anomalous Hall current in the active layer does not generate a drag current at all in the passive layer - both the longitudinal and the Hall components stemming from these topological terms vanish identically. Secondly, the remaining terms in the anomalous Hall drag current only depend on $M_\text{p}$, the magnetization of the passive layer. The dependence on $M_\text{p}$ is non-monotonic, with a peak at an intermediate value of $M_\text{p}$, which becomes pronounced at low densities. These results reveal the absence of a Hall drag \textit{per se} in anomalous Hall systems, and the fact that the measured Hall drag current is simply the anomalous Hall response to the longitudinal drag force. On a deeper level, given that the Berry curvature in solid state systems represents interband coherence, the results discussed here imply that interband coherence cannot be transferred between layers, at least not by electron-electron scattering. This is in sharp contrast to conventional Coulomb drag in ordinary Hall systems \cite{1995_Hall_drag,1999_Hall_drag, Hall_drag_Graphene, Song_Halldrag_2013, Sch_drag_prl_2013}. There the Hall current in the active layer is caused by the Lorentz force rather than topology or interband coherence and makes a significant contribution to the longitudinal and Hall drag response, the latter of which depends on the applied magnetic field.

Although our discussion is general and applies to materials with Rashba spin-orbit interactions even beyond topological insulators, the question of drag has a particular significance to the latter. Following their initial observation \citep{Urazhdin_PRB04, Hsieh_BiSb_QSHI_Nature08, Hsieh_BiSb_QmSpinTxtr_Science09, Xia_Bi2Se3_LargeGap_NP09, Hsieh_Bi2Te3_Sb2Te3_PRL09, Taskin_TI_eh_PRL11, Zhang_Bi2Se3_Film_Epitaxy_APL09, Wang_Bi2Te3_Ctrl_AM11, Benjamin_nature_2011}, improvements in TI growth have made them suitable for fundamental research \citep{Dohun_Surface_nphy_2012, Fuhrer_nature_2013, Brune_prx_2014, Hellerstedt_APL_2014}. A number of experiments have successfully identified signatures of the surface states in transport \citep{Benjamin_nature_2011, Fuhrer_nature_2013, Brune_prx_2014, Dohun_Surface_nphy_2012, Lucas_nl_tran_2014}. A current induced spin polarization also constitutes a signature of surface transport \citep{Burkov_TI_SpinCharge_PRL10, Culcer_TI_Kineq_PRB10} and was reported in recent experimental studies \citep{LiC.H_spin_polarization_nature2014, Jianshi_nl_spin_polarized_2014, Kastl_surface_tran_nature_2015}. However, the reliable and reproducible identification of the surface states in transport, which remains the key to topological insulators becoming technologically important, has remained elusive. At the same time the interplay of strong spin-orbit coupling and electron-electron interactions in topological insulators is at present not completely understood \citep{Culcer_TI_Int_PRB11, Yamaji-PRB-2011,Peters-PRL-2012,Yoshida-PRB-2012, Ostrovsky_TI_IntCrit_PRL10, WangCulcer_TI_Kondo_PRB2013, LiQiuzi_2013_prb,Hai-Zhou_Conductivity_2014_prl}, with transport experiments and theoretical mostly focused on single-particle effects \citep{Adam_2D_Tran_prb_2012, Qiuzi_2Dtran_phimu_2012_prb, JMShao-oscillation, Costache_prl_2014, Kozlov_prl_2014, Syers_SmB_prl_2015, Chang_TIF_ferromagnetism_AHE_AM2013, Kim-Hall2013, Jing_Shoucheng_QAHE_2014_prb, ShiJunren_QAHE_prl_2014, Dohun_Thermoelectric_nl_2014,Durst_2015,ZhangJinsong_prb_2015, He_Bi2Te3_Film_WAL_ImpEff_PRL11, Cheng_TI_STM_LL_PRL10, Tkachov_HgTe_WAL_PRB11, Competition_WL_WAL_2011}. As this work will show, plenty of scope exists for investigating interacting setups such as Coulomb drag geometries. 

The outline of this paper is as follows. In Sec.~\ref{sec:transport} we discuss a general theoretical framework for understanding Coulomb drag in systems in which spin/interband coherence is important, which requires the use of the spin density matrix. We discuss Coulomb drag for massless Dirac fermions in Sec.~\ref{sec:massless} and massive Dirac fermions in Sec.~\ref{sec:massive}. Generalisations of our picture beyond the prototype system of a topological insulator film are presented in Sec.~\ref{sec:beyond}. We end with a summary and conclusions. 

\section{Transport theory}
\label{sec:transport}

Quite generally, Dirac fermions in two dimensions are described by an effective Hamiltonian $H_{\text{D}} = A \,{\bm \sigma}\cdot({\bm k}\times\hat{{\bm z}}) + M\sigma_z$, with ${\bm \sigma}=(\sigma_x,\sigma_y,\sigma_z)$ the usual Pauli matrices, ${\bm k} = (k_x, k_y)$ the 2D wave vector, $A$ representing the Fermi velocity and $M$ a generic mass term. In the limit $M\rightarrow0$ the quasi-particle dispersion is linear.

We consider a TI film with both the top (active) and bottom (passive) surfaces magnetized either through doping with magnetic impurities or proximity coupling to ferromagnets. The magnetizations of the two surfaces are allowed to differ. Without loss of generality we assume (i) the carrier number density and hence the Fermi energy is the same in each layer and (ii) sidewall states do not participate in transport, an assumption that recent work has shown to be justified \cite{Tilahun_TIF_QHS_PRL2011}. The chemical potential lies in the surface conduction band of each layer. We require $\varepsilon_\text{F}\tau_l/\hbar\gg1$ in each layer $l \in \{\text{a} \equiv \text{active},\text{p} \equiv \text{passive}\}$, with $\varepsilon_\text{F}$ the Fermi energy located in the surface conduction bands and $\tau_l$ the momentum scattering time. The two-layer effective band Hamiltonian
\begin{equation}\label{band-H}
H_{0{\bm k}}=\tau_z\otimes h_{{\bm k}}+\text{diag}\{M_{\text{a}},-M_{\text{a}},M_{\text{p}},-M_{\text{p}}\},
\end{equation}
where the (Rashba) Hamiltonian of a single layer $h_{{\bm k}}= A\,{\bm \sigma}\cdot({\bm k}\times\hat{{\bm z}})\equiv -Ak{\bm \sigma}\cdot\hat{\bm \theta}$ with $\hat{\bm \theta}$ the tangential unit vector corresponding to ${\bm k}$. The Pauli matrix $\tau_z$ represents the layer degree of freedom. The eigenvalues of  Eq.~(\ref{band-H}) are $\varepsilon_{l\pm}={\pm}\sqrt{A^2k^2+M^2_{l}} \equiv \hbar\Omega_k^{(l)}$, the band index $s_{\bm k}=\pm$ with $+$ the conduction band and $-$ the valence band. 

The single-particle Hamiltonian $\hat{H}^{1e}=\hat{H}_0+\hat{H}_{E}+\hat{U}$, where $\hat{H}_0$ is the band Hamiltonian defined in Eq.~(\ref{band-H}), $\hat{H}_E=e(\hat{\bm E}\otimes\sigma_{0})\cdot\hat{\bm r}$ is the electrostatic potential due to the driving electric field with $\hat{\bm r}$ the position operator and $\hat{U}$ the disorder potential, which is assumed to be a scalar in spin space. Adding the two-particle interaction term we write the total Hamiltonian as $\hat{H}=\hat{H}^{1e}+\hat{V}^{ee}$ with the single-particle term expressed generically as $\hat{H}^{1e}\!=\!\sum_{\alpha\beta}H_{\alpha\beta}c^{\dag}_{\alpha}c_{\beta}$ and the Coulomb interaction term $\hat{V}^{ee}\!=\!\frac{1}{2}\sum_{\alpha\beta\gamma\delta}V^{ee}_{\alpha\beta\gamma\delta}c^\dag_\alpha c^\dag_\beta c_\gamma c_\delta$. The indices $\alpha\equiv{\bm k}s_{\bm k}l$ represent wave vector, band, and layer indices respectively. The matrix element $V^{ee}_{\alpha\beta\gamma\delta}$ in a generic basis $\{\phi_{\alpha}({\bm r})\}$ is given by
$V^{ee}_{\alpha\beta\gamma\delta}=\int \!d{\bm r}\int \!d{\bm r}' \ \phi^*_{\alpha}({\bm r})\phi^*_{\beta}({\bm r}')V^{ee}_{{\bm r}-{\bm r}'}\phi_{\delta}({\bm r})\phi_{\gamma}({\bm r}')$,
where $V^{ee}_{{\bm r}-{\bm r}'}$ is the Coulomb interaction in real space.

The dynamical screening function is found by solving the Dyson equation for the two-layer system in the random phase approximation (RPA) as in  Ref.~\citep{1995_Hall_drag}. The dynamically screened interlayer Coulomb interaction
\begin{equation}
V({\bm q},\omega)=\frac{v_q\text{e}^{-qd}}{\epsilon({\bm q},\omega)}.
\end{equation}
The dielectric function of the coupled layer system is
\begin{equation}\label{dielectric}
\arraycolsep 0.3ex
\begin{array}{rl}
\epsilon ({\bm q},\omega) & \dps =[1-v_q \Pi_\text{a} ({\bm q},\omega)][1-v_q \Pi_\text{p}({\bm q},\omega)]\\[3ex]
&\dps -[v_q\text{e}^{-qd}]^2\Pi_\text{a}({\bm q},\omega)\Pi_\text{p}({\bm q},\omega),
\ea
\end{equation}
in which the polarization function is obtained by summing the lowest bubble diagram and takes the form
\begin{equation}
\Pi_{l}({\bm q},\omega)=-\frac{1}{L^2}\sum_{{\bm k}ss'}\frac{(f^{(l)}_{0{\bm k},s}-f^{(l)}_{0{\bm k}',s'})}
{\varepsilon^{(l)}_{{\bm k},s}-\varepsilon^{(l)}_{{\bm k}',s'}+\hbar \omega+i0^{+}}F^{ss'}_{{\bm k}{\bm k}'},
\end{equation}
 with $f^{(l)}_{0{\bm k},s}\equiv f^{(l)}_0(\varepsilon_{{\bm k}s})$ the equilibrium Fermi distribution function and $F^{ss'}_{{\bm k}{\bm k}'}=\langle {\bm k},s,l|{\bm k}',s',l'\rangle \langle {\bm k}',s',l'|{\bm k},s,l\rangle$ is the wave function overlap. In magnetic TIs, due to the typical smallness of the gap opened by the magnetization, $F^{++}_{{\bm k}{\bm k}'} \approx \frac{1+\cos\gamma}{2}$, with $\gamma$ is the angle between the incident (${\bm k}$) and the outgoing wave vector (${\bm k}'$).  

Two parameters quantify the strength of interactions in topological insulator films and the limits of validity of the RPA. Firstly, we define an effective background dielectric constant $\epsilon_r$. The physics of films is determined by their thickness $d$ and the Fermi wave vector $k_F$ \cite{Tilahun_TIF_QHS_PRL2011}. We take a Bi$_2$Se$_3$ film as an example, with $\epsilon_{r,Bi2Se3} \approx 100$, grown on a semiconductor substrate with $\epsilon_{r,s} \approx 11$. For $k_Fd \gg 1$, the film is thick, and the two surfaces are independent \cite{AndoFowler_2D_RMP1982}. For the top surface, where one side is exposed to air, $\epsilon_{r, top} = (\epsilon_{r, Bi2Se3} + 1)/2 \approx 50$. For the bottom surface $\epsilon_{r, btm} = (\epsilon_{r, Bi2Se3} + \epsilon_{r, s})/2 \approx 55$. Both are independent of $d$. For $k_Fd \ll 1$, the film is ultrathin and can be approximated as a pure 2D system, with $\epsilon_r = (1 + \epsilon_{r, s})/2 \approx 6$, also independent of $d$. However, since the TI bulk cannot be eliminated, $\epsilon_r \approx 6$ is an ideal lower bound. In films studied here, thick enough that there is no interlayer tunnelling, implying $d > 5$ nm at the very least, $\epsilon_r$ has contributions from both the TI bulk and the semiconductor substrate, and for Bi$_2$Se$_3$ can range between 6 and 55. Two experiments have extracted $\epsilon_r \approx 30$ for relatively thick films of Bi$_2$Se$_3$ (10 nm $< d<$ 20 nm) \cite{Beidenkopf_NP11}. Hence, $\epsilon_r$ is treated as a phenomenological parameter to be measured separately for each film. In general $r_s \ll 1$ making the RPA approach applicable. The approximation $k_{\text{F}}d\gg1$ applies for TI film thicknesses up to $6\text{nm}$, with electron densities $n\thicksim10^{12}\text{cm}^{-2}$. This is in contrast to graphene, where $k_{\text{F}}d \ll1$ is usually satisfied. For other systems with small enough $k_{\text F}d$ and larger $r_s$, the interlayer Coulomb interaction is in general not small, and it remains to be determined whether it is necessary to take into account higher-order contributions in the Coulomb interaction.

\begin{figure}
\begin{center}
\includegraphics[width=1\columnwidth]{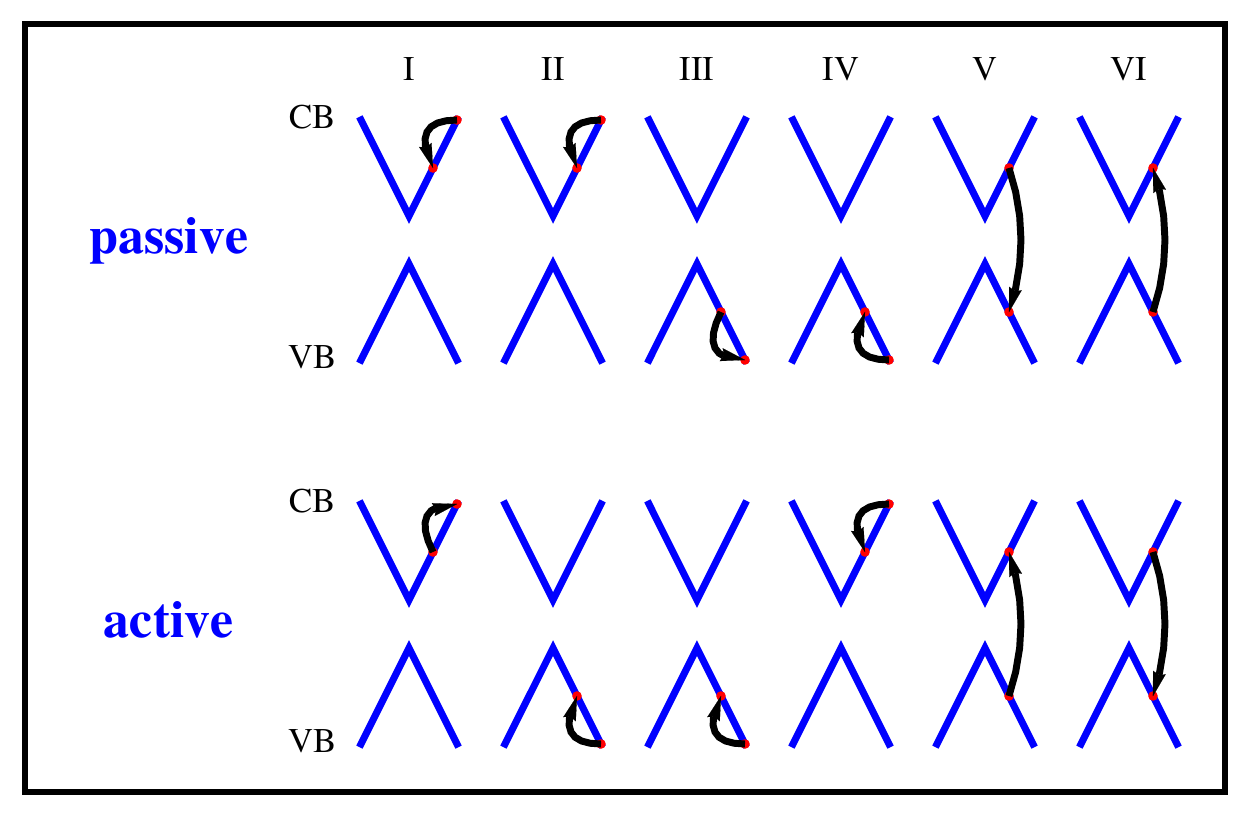}
\caption{\label{processes-6} Sketch of the interlayer electron-electron scattering processes that contribute to Coulomb drag. CB and VB represent the conduction and valence bands, respectively. Processes I-IV represent intraband scatterings, while interband scatterings are represented by V-VI.}
\end{center}
\end{figure}

The many-body density matrix $\hat{F}$ obeys the quantum Liouville equation \cite{Vasko}
\begin{equation}
\frac{\text{d}\hat{F}}{\text{d}t}+\frac{i}{\hbar}[\hat{H},\hat{F}]=0.
\end{equation}
The one-particle reduced density matrix is the trace $\text{tr}(c^\dag_\eta c_\xi \hat{F})\equiv\langle c^\dag_\eta c_\xi\rangle$. Its ${\bm k}$-diagonal part can be written as a $4 \times 4$ matrix in the joint spin/layer pseudospin subspace, and we refer to this matrix as $f_{\bm k}$. To second order in the electron-electron interaction, $f_{\bm k}$ satisfies \cite{Culcer_TI_Int_PRB11}
 \begin{equation}\label{kinetic_original_0}
 \frac{\text{d}f_{\bm k}}{\text{d}t} + \frac{i}{\hbar}[H^{1e}, f_{\bm k}] + \hat{J}_{ee}(f_{\bm k}) =0.
 \end{equation}
The term $\hat{J}_{ee}(f_{\bm k})$ in Eq.~(\ref{kinetic_original_0}) represents intralayer and interlayer electron-electron scattering operator. Since the intralayer electron-electron scattering does not contribute to the drag current, we concentrate on the interlayer term which is the matrix form of operator $\hat{J}_{ee}(f_{\bm k})$,
\begin{equation}\label{electron-electron-m}
\ba
&\dps J_{ee}^i (f_{\bm k})_{s_{\bm k}s'_{\bm k}} = \frac{\pi}{\hbar L^4}\sum_{{\bm k}_1{\bm k}'{{\bm k}'_1}}|v^{(\text{pa})}_{|{\bm k}-{\bm k}_1|}|^2\delta_{{\bm k}+{\bm k}',{\bm k}_1+{\bm k}'_1}\big\{\\[2ex]
&\dps \sum_{i=1}^{4}\!P^{(i)}_{s_{\bm k}s_{{\bm k}_1}s'_{\bm k}}A^{(i)}_{s_{\bm k'}s_{{\bm k}'_1}}\delta[\varepsilon^{(\text{p})}_{k_1,s_{{\bm k}_1}}\!-\varepsilon^{(\text{p})}_{k,s'_{\bm k}}\!+\varepsilon^{(\text{a})}_{k'_1,s_{{\bm k}'_1}}\!-\varepsilon^{(\text{a})}_{k',s_{\bm k'}}]+\\[2ex]
&\dps \sum_{i=5}^8\!P^{(i)}_{s_{\bm k}s_{{\bm k}_1}s'_{\bm k}}A^{(i)}_{s_{\bm k'}s_{{\bm k}'_1}}\delta[\varepsilon^{(\text{p})}_{k_1,s_{{\bm k}_1}}\!-\varepsilon^{(\text{p})}_{k,s_{\bm k}}\!+\varepsilon^{(\text{a})}_{k'_1,s_{{\bm k}'_1}}\!-\varepsilon^{(\text{a})}_{k',s_{\bm k'}}]\big\},
\ea
\end{equation}
where $L^2$ is the area of the 2D system and the quantities $A^{(i)}_{s_{\bm k'}s_{{\bm k}'_1}}$ and $P^{(i)}_{s_{\bm k}s_{{\bm k}_1}s'_{\bm k}}$, given explicitly in the Table~{\color{blue} 1}, are functions of the occupations of the active and passive layers respectively. The interlayer momentum transfer ${\bm q}={\bm k}-{\bm k}_1={\bm k}'_1-{\bm k}'$, and $v^{(\text{pa})}_{|{\bm k}-{\bm k}_1|}$ is the interlayer Coulomb interaction. The interlayer electron-electron scattering processes that contribute to Coulomb drag are shown in Fig.~\ref{processes-6}.
\begin{table}
\centering
\caption{$A^{(i)}_{s_{\bm k'}s_{{\bm k}'_1}}$ and $P^{(i)}_{s_{\bm k}s_{{\bm k}_1}s'_{\bm k}}$}
\begin{tabular}{|c|}
\hline
$P^{(1)}_{s_{\bm k}s_{{\bm k}_1}s'_{\bm k}}=-\langle s_{\bm k}|s_{{\bm k}_1}\rangle\left[1-f^{(\text{p})}_{0{\bm k}_1,s_{{\bm k}_1}}\right]\langle s_{{\bm k}_1} | s'_{\bm k} \rangle f^{(\text{p})}_{0{\bm k},s'_{\bm k}}$ \\
\hline
$P^{(2)}_{s_{\bm k}s_{{\bm k}_1}s'_{\bm k}}=\langle s_{\bm k}|s_{{\bm k}_1}\rangle f^{(\text{p})}_{0{\bm k}_1,s_{{\bm k}_1}}\langle s_{{\bm k}_1} | s'_{\bm k} \rangle \left[1-f^{(\text{p})}_{0{\bm k},s'_{\bm k}}\right]$ \\
\hline
$P^{(3)}_{s_{\bm k}s_{{\bm k}_1}s'_{\bm k}}=P^{(8)}_{s_{\bm k}s_{{\bm k}_1}s'_{\bm k}}=-\langle s_{\bm k}|s_{{\bm k}_1}\rangle f^{(\text{p})}_{0{\bm k}_1,s_{{\bm k}_1}}\langle s_{{\bm k}_1} | s'_{\bm k} \rangle$ \\
\hline
$P^{(4)}_{s_{\bm k}s_{{\bm k}_1}s'_{\bm k}}=\langle s_{\bm k}|s_{{\bm k}_1}\rangle \langle s_{{\bm k}_1} | s'_{\bm k} \rangle f^{(\text{p})}_{0{\bm k},s'_{\bm k}}$ \\
\hline
$P^{(5)}_{s_{\bm k}s_{{\bm k}_1}s'_{\bm k}}=\left[1-f^{(\text{p})}_{0{\bm k},s_{\bm k}}\right]\langle s_{\bm k}|s_{{\bm k}_1}\rangle f^{(\text{p})}_{0{\bm k}_1,s_{{\bm k}_1}}\langle s_{{\bm k}_1} | s'_{\bm k} \rangle$ \\
\hline
$P^{(6)}_{s_{\bm k}s_{{\bm k}_1}s'_{\bm k}}=-f^{(\text{p})}_{0{\bm k},s_{\bm k}}\langle s_{\bm k}|s_{{\bm k}_1}\rangle\left[1-f^{(\text{p})}_{0{\bm k}_1,s_{{\bm k}_1}}\right]\langle s_{{\bm k}_1} | s'_{\bm k} \rangle$ \\
\hline
$P^{(7)}_{s_{\bm k}s_{{\bm k}_1}s'_{\bm k}}=f^{(\text{p})}_{0{\bm k},s_{\bm k}}\langle s_{\bm k}|s_{{\bm k}_1}\rangle \langle s_{{\bm k}_1} | s'_{\bm k} \rangle$ \\
\hline
\hline
 $A^{(1)}_{s_{\bm k'}s_{{\bm k}'_1}}=\sum_{\varsigma_{\bm k'}}=\langle s_{{\bm k}'} |s_{{\bm k}'_1}\rangle\langle s_{{\bm k}'_1}| \varsigma_{{\bm k}'} \rangle f^{(\text{a})}_{{\bm k}',\varsigma_{\bm k'}s_{\bm k'}}$ \\
 \hline
 $A^{(2)}_{s_{\bm k'}s_{{\bm k}'_1}}=\sum_{\varsigma_{{\bm k}'_1}}=\langle s_{{\bm k}'} |s_{{\bm k}'_1}\rangle f^{(\text{a})}_{{\bm k}'_1,s_{{\bm k}'_1}\varsigma_{{\bm k}'_1}}\langle \varsigma_{{\bm k}'_1}| s_{{\bm k}'}\rangle$  \\
 \hline
 $A^{(3)}_{s_{\bm k'}s_{{\bm k}'_1}\varsigma_{{\bm k}'_1}\varsigma_{\bm k'}}=\langle s_{{\bm k}'} |s_{{\bm k}'_1}\rangle f^{(\text{a})}_{{\bm k}'_1,s_{{\bm k}'_1}\varsigma_{{\bm k}'_1}}\langle \varsigma_{{\bm k}'_1}| \varsigma_{{\bm k}'}\rangle f^{(\text{a})}_{{\bm k}',\varsigma_{\bm k'}s_{\bm k'}}$  \\
 \hline
  $A^{(4)}_{s_{\bm k'}s_{{\bm k}'_1}\varsigma_{{\bm k}'_1}\varsigma_{\bm k'}}=\langle s_{{\bm k}'} |s_{{\bm k}'_1}\rangle f^{(\text{a})}_{{\bm k}'_1,s_{{\bm k}'_1}\varsigma_{{\bm k}'_1}}\langle \varsigma_{{\bm k}'_1}| \varsigma_{{\bm k}'}\rangle f^{(\text{a})}_{{\bm k}',\varsigma_{\bm k'}s_{\bm k'}}$  \\
 \hline
 $A^{(5)}_{s_{\bm k'}s_{{\bm k}'_1}}=\sum_{\varsigma_{{\bm k}'_1}}=\langle s_{{\bm k}'}|\varsigma_{{\bm k}'_1}\rangle f^{(\text{a})}_{{\bm k}'_1,\varsigma_{{\bm k}'_1}s_{{\bm k}'_1}}\langle s_{{\bm k}'_1}|s_{{\bm k}'}\rangle$ \\
 \hline $A^{(6)}_{s_{\bm k'}s_{{\bm k}'_1}}=\sum_{\varsigma_{\bm k'}}\langle \varsigma_{{\bm k}'}|s_{{\bm k}'_1}\rangle\langle s_{{\bm k}'_1}|s_{{\bm k}'}\rangle f^{(\text{a})}_{{\bm k}',s_{\bm k'}\varsigma_{\bm k'}}$  \\
 \hline $A^{(7)}_{s_{\bm k'}s_{{\bm k}'_1}\varsigma_{{\bm k}'_1}\varsigma_{\bm k'}}=\langle s_{{\bm k}'_1}|s_{{\bm k}'}\rangle f^{(\text{a})}_{{\bm k}',s_{\bm k'}\varsigma_{\bm k'}} \langle \varsigma_{{\bm k}'}|\varsigma_{{\bm k}'_1}\rangle f^{(\text{a})}_{{\bm k}'_1,\varsigma_{{\bm k}'_1}s_{{\bm k}'_1}}$ \\
  \hline $A^{(8)}_{s_{\bm k'}s_{{\bm k}'_1}\varsigma_{{\bm k}'_1}\varsigma_{\bm k'}}=\langle s_{{\bm k}'_1}|s_{{\bm k}'}\rangle f^{(\text{a})}_{{\bm k}',s_{\bm k'}\varsigma_{\bm k'}} \langle \varsigma_{{\bm k}'}|\varsigma_{{\bm k}'_1}\rangle f^{(\text{a})}_{{\bm k}'_1,\varsigma_{{\bm k}'_1}s_{{\bm k}'_1}}$ \\
\hline
\end{tabular}
\end{table}

At this stage one may include explicitly disorder and driving electric field in the one particle Hamiltonian and write $\hat{H}^{1e}=\hat{H}_0+\hat{H}_{E}+\hat{U}$, where $\hat{H}_0$ is the band Hamiltonian of TIs, $\hat{H}_E=e\hat{\bm E}\cdot\hat{\bm r}$ is the electrostatic potential due to the driving electric flied with $\hat{\bm r}$ is a position operator and $\hat{U}$ is the disorder potential. Referring to the layer index $^{(l)}$, we write $f^{(l)}_{\bm k}=n^{(l)}_{\bm k}\mathbb{1}+S^{(l)}_{\bm k}$, with $S^{(l)}_{\bm k}$ a $2\times2$ Hermitian matrix which can be decomposed in terms of the Pauli spin matrices. The single-particle density matrix is diagonal in the layer index $f_{\bm k} = {\rm{diag}} \, (f_{\bm k}^{(\text{a})}, f_{\bm k}^{(\text{p})})$. The scattering term $J^i_{ee} (f_{\bm k})_{s_{\bm k}s'_{\bm k}}$ represents interlayer coherence: in the active layer we solve for the non-equilibrium correction to $f_{\bm k}^{(\text{a})}$, feed it into $J^i_{ee}(f_{\bm k})_{s_{\bm k}s'_{\bm k}}$, and the result is an effective driving term for the passive layer. This then enables us to solve for the non-equilibrium to correction $f_{\bm k}^{(\text{p})}$ in response to this effective driving term. The kinetic equation of the two-layer system
\begin{subequations}\label{eqab}
 \begin{equation}\label{active}\vspace{-0.5cm}
  \frac{\text{d}f^{(\text{a})}_{\bm k}}{\text{d}t}+\frac{i}{\hbar}[H^{(\text{a})}_{0{\bm k}},f^{(\text{a})}_{\bm k}]+\hat{J}_{0}(f^{(\text{a})}_{\bm k}) = \frac{e{\bm E}}{\hbar}\cdot \pd{f^{(\text{a})}_{0{\bm k}}}{{\bm k}},  
 \end{equation}
\begin{equation}\label{passive}
\frac{\text{d}f^{(\text{p})}_{\bm k}}{\text{d}t}+\frac{i}{\hbar}[H^{(\text{p})}_{0{\bm k}},f^{(\text{p})}_{\bm k}]+\hat{J}_{0}(f^{(\text{p})}_{\bm k}) = J^i_{ee} (f_{\bm k})_{s_{\bm k}s'_{\bm k}} ,
\end{equation}
\end{subequations}
with $H^{(l)}_{0{\bm k}}=h^{(l)}_{\bm k}+M_l \sigma_z$ the band Hamiltonian and $f^{(l)}_{0{\bm k}}$ the equilibrium density matrix of each magnetic layer. The electron-impurity scattering integral is $\hat{J}_{0}(f^{(l)}_{\bm k})~=~\bigg\langle\int^\infty_0\frac{\text{d}t'}{\hbar^2}[\hat{U},e^{-i\hat{H}t'/\hbar}[\hat{U},\hat{f}]
e^{i\hat{H}t'/\hbar}]\bigg\rangle_{{\bm k}{\bm k}}$, where $\langle\rangle$ denotes the average over impurity configurations. In EQ.~(\ref{passive}), $J^i_{ee} (f_{\bm k})_{s_{\bm k}s'_{\bm k}}$ can be rewritten as
\begin{equation}\label{Jmatrix}
\ba
  J^i_{ee}(f_{\bm k})_{s_{\bm k}s'_{\bm k}}&\dps
 =\!\frac{J^i_{ee}(f_{\bm k})_{++}+J^i_{ee}(f_{\bm k})_{--}}{2}\sigma_0\\[3ex]
 &\dps +\frac{J^i_{ee}(f_{\bm k})_{++}-J^i_{ee}(f_{\bm k})_{--}}{2}\sigma_z\\[3ex]
 &\dps +\frac{J^i_{ee}(f_{\bm k})_{-+}+J^i_{ee}(f_{\bm k})_{+-}}{2}\sigma_x\\[3ex]
&\dps  +\frac{J^i_{ee}(f_{\bm k})_{-+}-J^i_{ee}(f_{\bm k})_{+-}}{2i}\sigma_y,
\ea
\end{equation}
with $\sigma_0$ is $2\times2$ unit matrix and $\sigma_{x,y,z}$ are Pauli matrix. We should transform Eq.~\ref{Jmatrix}  from eigenstates representation (ER) to that of of $\sigma_z$ (ordinary representation, OR) with 
\begin{equation}\label{T}
\text{T}
=\left(
\begin{array}{cc}
\dps-i\text{e}^{i\theta_{\bm k}}\sqrt{\frac{1+b_k}{2}}  & \quad \dps\frac{a_k}{\sqrt{2(1+b_k)}}  \\[3ex]
\dps i\text{e}^{i\theta_{\bm k}}\sqrt{\frac{1-b_k}{2}}  & \quad \dps\frac{a_k}{\sqrt{2(1-b_k)}}
\end{array}
\right),
\end{equation}
and for any $2\times2$ matrix $\mathcal{M}$, $  \mathcal{M}^{\text{OR}}=\text{T}^{\dag}\cdot\mathcal{M}^{\text{ER}}\cdot\text{T}$. We have
$\sigma_z \to {\bm \sigma}\cdot\hat{{\bm \Omega}}_{\bm k}, \sigma_x \to -{\bm \sigma}\cdot\hat{{\bm z}}_{\text{eff}},
\sigma_y \to {\bm \sigma}\cdot\hat{{\bm k}}_{\text{eff}}$. In $\sigma_z$ representation, according to the contributions to various drag currents, the electron-electron scattering term is divided into
\begin{equation}
\ba
 J^{i}_{ee}(f_{\bm k})_\text{ll,lh}&\dps =-\!\frac{J^i_{ee}(f_{\bm k})_{++}+J^i_{ee}(f_{\bm k})_{--}}{2}\sigma_0\\[3ex] 
&\dps -\frac{J^i_{ee}(f_{\bm k})_{++}-J^i_{ee}(f_{\bm k})_{--}}{2}{\bm \sigma}\cdot\hat{\bm \Omega}_{\bm k},\\[3ex]
 J^{i}_{ee}(f_{\bm k})_\text{hl,hh} &\dps =\frac{J^i_{ee}(f_{\bm k})_{-+}\!+\!J^i_{ee}(f_{\bm k})_{+-}}{2}{\bm \sigma}\cdot\hat{\bm z}_{\text{eff}}\\[3ex]
&\dps -\frac{J^i_{ee}(f_{\bm k})_{-+}\!-\!J^i_{ee}(f_{\bm k})_{+-}}{2i}{\bm \sigma}\cdot\hat{\bm k}_{\text{eff}},
\ea
\end{equation}
where ll,lh,hl,hh represent contributions to the four drag current components in Fig.~\ref{4-currents}.

The above encapsulates the philosophy or our approach: to begin with, we consider an external electric field applied to the active layer, we solve for the non-equilibrium density matrix in the active layer without any reference to electron-electron scattering, and we feed the solution for the active-layer density matrix into the interlayer electron-electron scattering integral. The result of this is an electric-field dependent term that acts as a new driving term (i.e. \textit{the drag force}) for the passive layer. We then solve for the non-equilibrium density matrix in the passive layer with this driving term. The solution thus obtained represents the non-equilibrium density matrix in the passive layer, which may be used to calculate expectation values. Specifically, its trace with the current operator yields the drag current. 

\section{Massless Dirac fermions}
\label{sec:massless}

Physically the drag current is a result of the rectification by the passive layer of the fluctuating electric field created by the active layer \citep{1995_Hall_drag}. At low temperatures the predominant contribution to drag is due to intraband processes near the Fermi surface. When the Fermi level is finite, i.e. above the Dirac point, electrons take more energy to transition from the valence band to the conduction band than to transition within the conduction band or valence band, and with a small excitation energy the channel involving interband transitions becomes inaccessible \citep{Tse2007}. The small interlayer momentum transfer and excitation energy is the dominant part of the spectrum contributing to the drag current. 

\begin{figure}[tbp]
\begin{center}
\includegraphics[width=1\columnwidth]{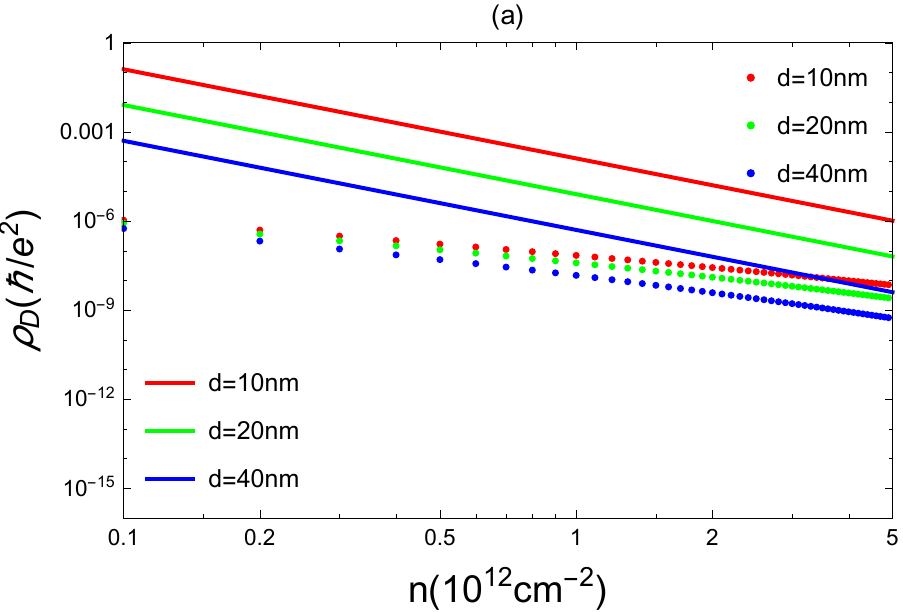}\\
\includegraphics[width=1\columnwidth]{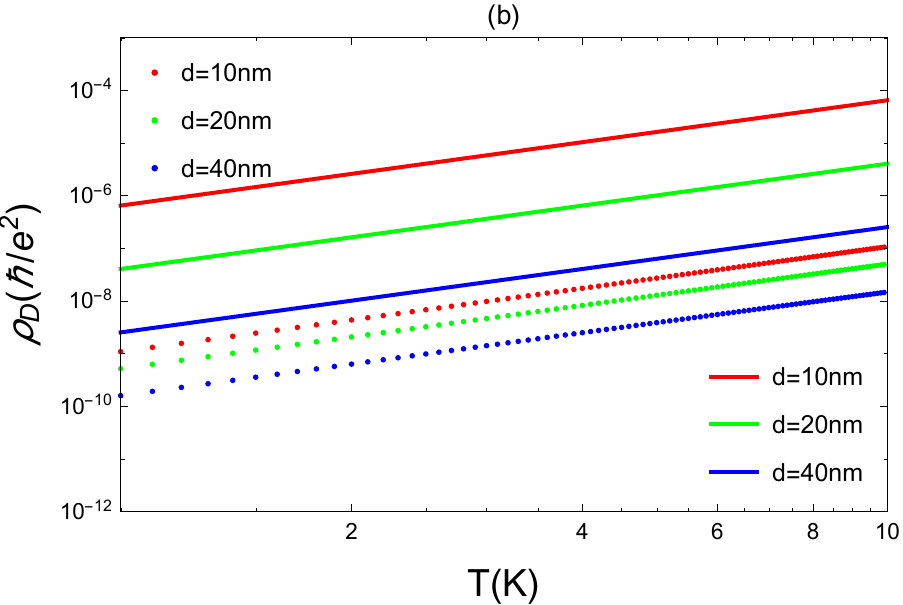}\\
\includegraphics[width=1\columnwidth]{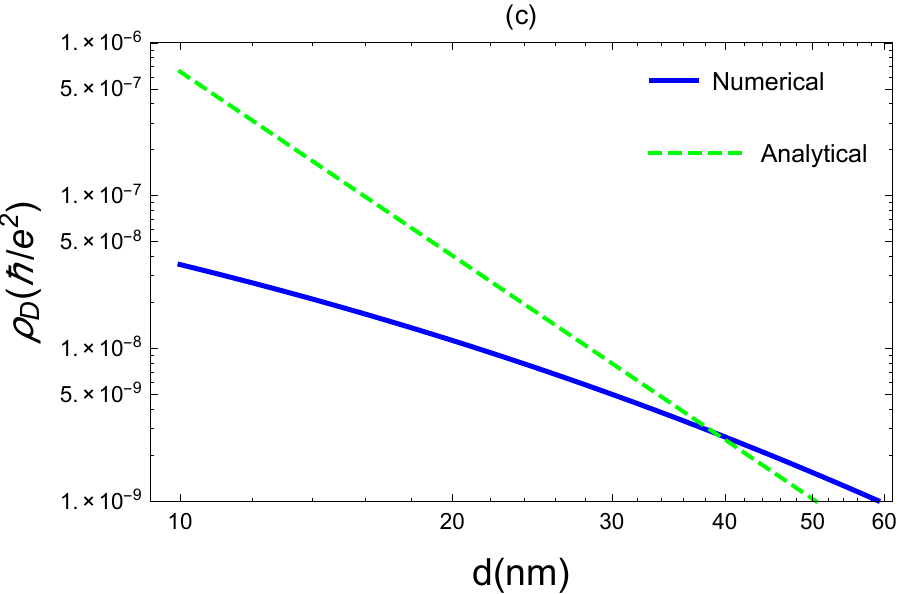}
\caption{\label{xxD} Behavior of drag resistivity $\rho_{\text{D}}$ as a function of electron concentration $n$, temperature $T$, layer separation $d$: (a) is electron density dependence of $\rho_\mathrm{D}$ at $T=5~\mathrm{K}$; (b) is temperature dependence of $\rho_\mathrm{D}$ with $k_\mathrm{F_{\mathrm{a}}}=k_\mathrm{F_{\mathrm{p}}}=0.5~\mathrm{nm}$; (c) is layer separation $d$ dependence of $\rho_\mathrm{D}$ at $T=5\text{K}$ and $k_\mathrm{F_{\mathrm{a}}}=k_\mathrm{F_{\mathrm{p}}}=0.5~\mathrm{nm}$.   Dielectric constant $\epsilon_r=50$.  Real and dotted (dashed) lines represent the numerical  and analytical  respectively.}
\end{center}
\end{figure}

The longitudinal drag conductivity $\sigma_D$ is
\begin{equation}\label{cll}
\sigma_\text{D}\!=\!\frac{e^2\beta}{16\pi}\!\sum_{\bm q}\!\int\!\text{d}\omega\frac{|v_{q}^{(\text{pa})}|^2\text{Im}\chi_\text{a}({\bm q},
\omega)\text{Im}{\chi}_\text{p}({\bm q},
\omega)}{\sinh^2 \frac{\beta\hbar\omega}{2}},
\end{equation}
where $\text{Im}\chi_l({\bm q},\omega)$ is the imaginary part of nonlinear susceptibility and $\beta\!=\!\frac{1}{k_{\text{B}}T}$. Using standard approximations we can express the drag resistivity analytically as
\begin{equation}\label{ll-result}
\rho_\text{D} \approx-\frac{\sigma_{\text{D}}}{\sigma_{\text{a}}\sigma_{\text{p}}} 
=-\frac{\hbar}{e^2}\frac{\zeta(3)}{16\pi}\frac{(k_\text{B}T)^2}{A^2r^2_sn^{\frac{3}{2}}_\text{a}n^{\frac{3}{2}}_\text{p}d^4},
\end{equation}
where $\sigma_{\text{a,p}}$ is the longitudinal conductivity of each layer.

In Fig.~\ref{xxD}(a) we present numerical results for the dependence of the Coulomb drag resistivity on the electron number density. The drag resistivity displays a $\frac{1}{n^{\alpha}_{\text{a}}n^{\alpha}_{\text{p}}}$ dependence where $\alpha$ is the power exponent of the density $n_{\text{a,p}}$, with $\alpha<1.5$ for $d=10,20,40~\text{nm}$. With increasing electron density $n$ the coefficient $\alpha$ approaches $1.5$. The fact that the exact numerical results shown in Fig.~\ref{xxD}(a) disagree more strongly with the analytical results for smaller values of $n$ and $d$ is understandable, since  Eq. (\ref{ll-result}) applies only in the $k_\text{F} d\gg1$ limit with $k_\text{F}=\sqrt{4\pi n}$. This trend of an increasing quantitative failure of the asymptotic analytical drag formula for small $k_\text{F}d$ has also been noted in graphene \citep{Tse2007,Hwang2011,M.Carrega2012} and 2DEG systems \citep{1995_Hall_drag,1999_Hall_drag}. The analytical result becomes more accurate with increasing $k_\text{F}d$.

In Fig.~\ref{xxD}(b) we show the Coulomb drag resistivity as a function of temperature $T$ for three different thicknesses $d=10,20,40~\text{nm}$. The overall temperature dependence of the drag resistivity increases nearly quadratically and there is no logarithmic correction due to the absence of backscattering in TIs. The $T^2$ dependence stems from the allowed phase space where electron-electron scattering occurs at low temperature, and is expected for any interaction strength between the top and bottom layers of TIs as Fig.~\ref{xxD}(b) shows, provided that the carriers can be described using a Fermi liquid picture. In addition, in TIs the acoustic phonon velocity is smaller than in graphene. These facts make the contribution of electron-phonon scattering processes to the resistivity much more important in the surface of 3DTIs than in graphene. For the surfaces of 3DTIs the effect of electron-phonon scattering events becomes important already for $T$ as low as $10\text{K}$. For this reason we consider temperatures up to $10\text{K}$ in our numerical calculations. In graphene this effect becomes relevant only beyond $T\gtrsim200\text{K}$. It is also evident that Eq. (\ref{ll-result}) becomes increasingly accurate and approaches the numerical results as the layer separation $d$ increases.

The behavior of the drag resistivity $\rho_{\text{D}}$ as a function of layer separation $d$ is shown in Fig.~\ref{xxD}(c) with $T=5\text{K}$ and $k_\mathrm{F_{\mathrm{a}}}=k_\mathrm{F_{\mathrm{p}}}=0.5~\mathrm{nm}$. The trend of the exact numerical results changes more slowly, a fact that is also embodied in Fig.~\ref{xxD}(a). Interestingly, in drag experiments on graphene \citep{Gorbachevi_nature_2012}, the $d$ dependence of the drag resistivity is much slower than the $1/d^4$ expected in the weakly interacting regime, varying approximately as $1/d^2$ for $d>4\text{nm}$, which is comparable to TIs.

\section{Massive Dirac fermions}
\label{sec:massive}

In a magnetic system the current in the active layer has a longitudinal and a Hall component, the latter being a result of the anomalous Hall effect. The principal mechanisms behind the anomalous Hall effect are: \cite{Nagaosa-AHE-2010, Culcer_TI_AHE_PRB11} (i) a topological mechanism related to spin precession under the combined action of spin-orbit coupling and the external electric field, which is strongly renormalized by scalar scattering (ii) three spin-dependent scattering mechanisms, which were shown to be of secondary importance in TIs, and will not be considered further. The drag current correspondingly has a longitudinal component and a Hall component, which can be measured separately. If we regard the current in the active layer as giving rise to an effective drag force, it is natural to ask which components of this drag force make the most sizable contribution to the drag current. Indeed, if we focus on the Hall component of the drag current we identify two potential contributions having physically distinct origins. Firstly, the longitudinal current in the active layer can give rise to a \textit{transverse drag force} in the passive layer, resulting in a Hall current. Secondly, the Hall current in the active layer, stemming from a topological mechanism, can itself directly drag a Hall current in the passive layer, in what may be termed \textit{direct Hall drag}.

We divide the drag current into four contributions, corresponding to the picture presented in Fig.~\ref{4-currents} in the introduction. The electric field $\parallel\hat{\bm x}$ gives rise to a longitudinal (${\bm j}_x$) and an anomalous Hall (${\bm j}_{\text{AHE}}$) current in the active layer. In the passive layer there are four contributions to the drag currents: ${\bm j}^{\text{ll}}$ is the longitudinal current dragged by ${\bm j}_x$; ${\bm j}^{\text{lh}}$ is the transverse current dragged by ${\bm j}_{\text{AHE}}$; ${\bm j}^{\text{hl}}$ is the anomalous Hall current generated by ${\bm j}^{\text{ll}}$; while ${\bm j}^{\text{hh}}$ is the anomalous Hall current generated by ${\bm j}^{\text{lh}}$, and it flows longitudinally. In the regime $M_{\text{a,p}} \ll \varepsilon_\text{F}$, which is applicable to all samples studied experimentally, the longitudinal drag conductivity $\sigma^{xx}_D$ does not depend on the magnetization of either layer, and has the same form as in Eq. (\ref{ll-result}). Next, the transverse drag ${\bm j}^{\mathrm{hl}}$, which is the anomalous Hall current due to ${\bm j}^{\mathrm{ll}}$ yields $\rho^{yx}_\text{D}\!\approx\!\frac{\sigma^{yx}_{\text{pa}}}{\sigma_{\text{a}}\sigma_{\text{p}}}$. Finally ${\bm j}^{\text{lh}} = {\bm j}^{\text{hh}} =0$. Referring to Fig.~\ref{4-currents} this implies that the anomalous Hall current in the (doped) active layer does not generate a corresponding drag current in the passive layer. 

The physics can be understood in two steps. Firstly, an external electric field drives the electrons in the active layer longitudinally, and these in turn exert a longitudinal drag force on the passive-layer electrons. The drag force  acts as an effective longitudinal driving term for the passive-layer electrons. The anomalous Hall component of the drag current, which at low temperatures can be sizeable compared to the longitudinal component, represents the transverse response of the passive layer to this effective longitudinal driving force, and depends on the passive-layer magnetization $M_\text{p}$. The electric field also generates an anomalous Hall current in the active layer. This response is dominated by topological terms of the order of the conductivity quantum, which represents a re-arrangement of charge carriers among spin-momentum locked energy states. Physically this is because Coulomb drag occurs as a result of the interaction between the charge densities in different layers, whereas the anomalous Hall current flowing in the active layer is not associated with a change in the charge density: it does not arise from a shift in the Fermi surface, but from the Berry phase acquired through the rearrangement of carriers among spin-momentum locked states. This implies that the anomalous Hall drag current is quite generally independent of the active-layer magnetization $M_\text{a}$. This can be easily verified experimentally.

We analyze the parameter dependence of $\rho^{yx}_\text{D}$ in Fig.~\ref{yxM} and Fig.~\ref{yx3}. The relationship between $\rho^{yx}_\text{D}$ and the magnetization of the passive layer is illustrated in Fig.~\ref{yxM}. There is an upward trend at small magnetizations followed by a relatively slow decrease at larger values of $M_{\text{p}}$. The trend can be understood as follows. For $M_{\text{p}}\ll Ak_{\text{F}}$ one may expand ${\bm j}^{\text{hl}}$ in $M_{\text{p}}$, which reveals that the current increases nearly linearly with $M_{\text{p}}$. In the opposite limit in which $M_{\text{p}}\gg Ak_{\text{F}}$, \cite{AHE_vertex_PRL_2006} the anomalous Hall current vanishes, since the effect of spin-orbit coupling (chirality), which scales with $k_{\text{F}}$, becomes negligible. In fact at large $M_{\text{p}}$ one may expand the band energies in $Ak$, with the leading $k$-dependent term scaling as $k^2$, which suggests the system in this limit behaves as a regular, non-magnetic 2DEG. This explains the slow downward trend with increasing $M_{\text{p}}$, and hence the presence of the peak as a function of $M_{\text{p}}$. At larger electron densities the peak occurs at stronger values of the magnetization, since higher electron densities imply higher values of $Ak_\text{F}$, increasing the effect of chirality at the expense of the magnetization. TI magnetizations have been measured by superconducting quantum interference device magnetometers.\cite{Hor_DopedTI_FM_PRB10,Jinsong-Zhang-Yayu-Wang-2013-science} Because the momentum scattering time is in principle known from the longitudinal conductivity of each layer, \cite{Cacho_TI_tau,Peng_natcom_2016_tau} the trend exhibited by $\rho^{yx}_\text{D}$ as a function of the magnetization $M_{\text{p}}$ can be verified experimentally.

\begin{figure}\label{yxM}
\begin{center}
\includegraphics[width=1\columnwidth]{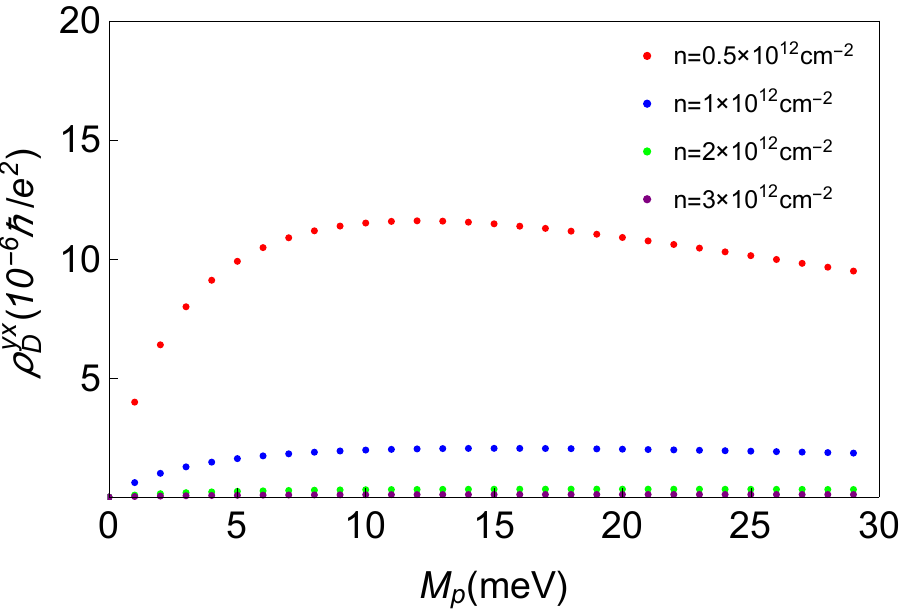}
\caption{\label{yxM} Magnetization dependence of
$\rho^{yx}_\mathrm{D}$ with $T=5~\text{K}$, $d=10~\text{nm}$, dielectric constant $\epsilon_r=20$, $A=4.1~\text{eV\AA}$ and transport time $\tau\approx0.1~\mathrm{ps}$.}
\end{center}
\end{figure}

We examine the dependence of the anomalous Hall drag resistivity on additional experimentally measurable parameters. In Fig.~\ref{yx3}(a) the electron density dependence of $\rho^{yx}_{\text{D}}$ for interlayer separations $d=10,20,40 ~\text{nm}$ is shown. Compared with the longitudinal drag resistivity, the anomalous Hall drag resistivity has a weaker dependence on electron density. As compared with the longitudinal drag resistivity, the group velocity appearing in the susceptibility of the passive layer is replaced by the Berry curvature, leading to a weaker density dependence, yet no topological contribution. Next, Fig.~\ref{yx3}(b) illustrates the temperature dependence of $\rho^{yx}_{\text{D}}$ for separations $d=10,20,40 ~\text{nm}$. We find that, much like $\rho^{xx}_{\text{D}}$, $\rho^{yx}_{\text{D}}$ also increases nearly quadratically with temperature. The $T^2$ dependence stems from the allowed phase space for electron-electron scattering at low temperature, and is expected for any interaction strength between the top and bottom layers of TIs, provided the carriers can be described using a Fermi liquid picture. Moreover, due to the absence of backscattering in TIs there is no correction logarithmic in temperature. Fig.~\ref{yx3}(c) presents the layer separation dependence of $\rho^{yx}_{\text{D}}$ for  $k_\text{F}= 0.4, 0.5, 0.6 \text{nm}^{-1}$. 

\begin{figure}\label{yx3}
\begin{center}
\includegraphics[width=1\columnwidth]{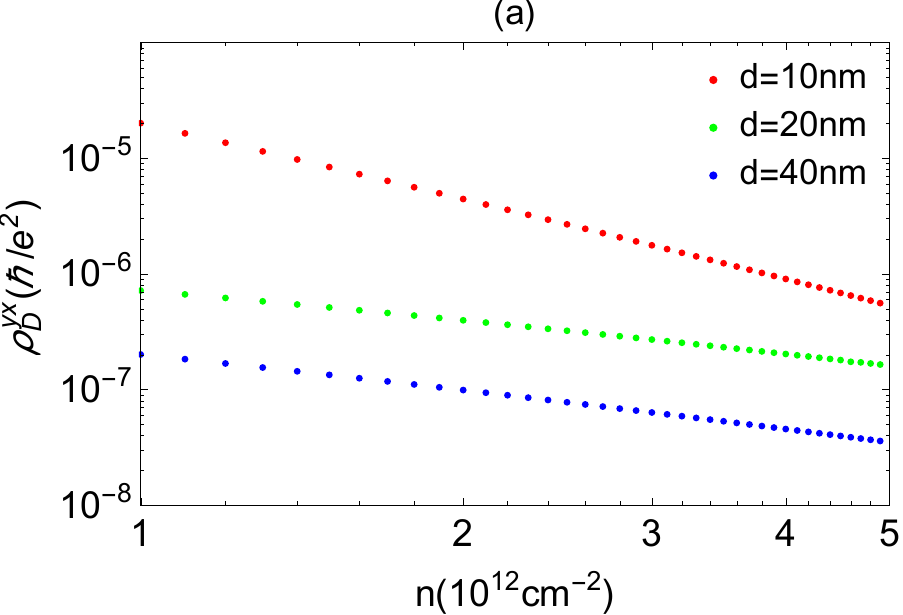}
\includegraphics[width=1\columnwidth]{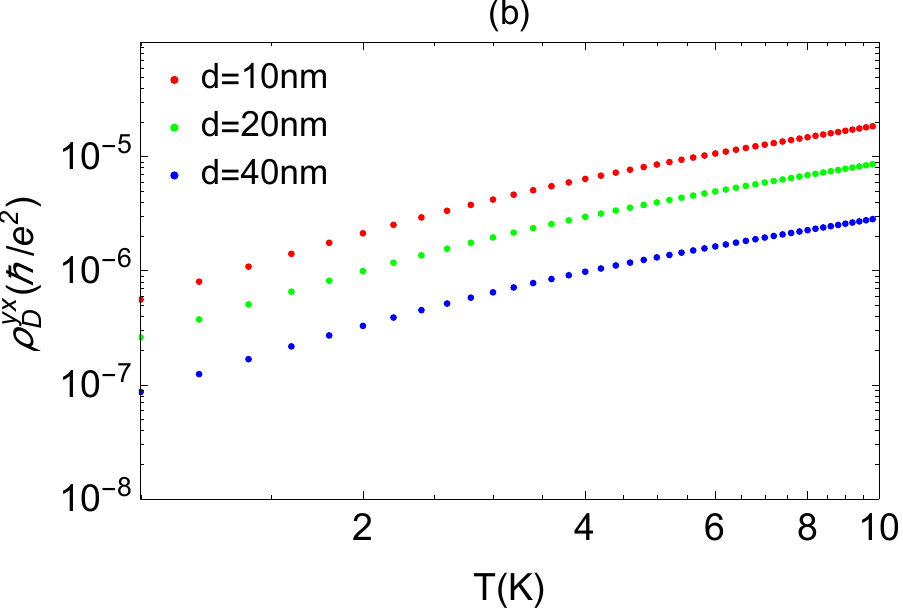}
\includegraphics[width=1\columnwidth]{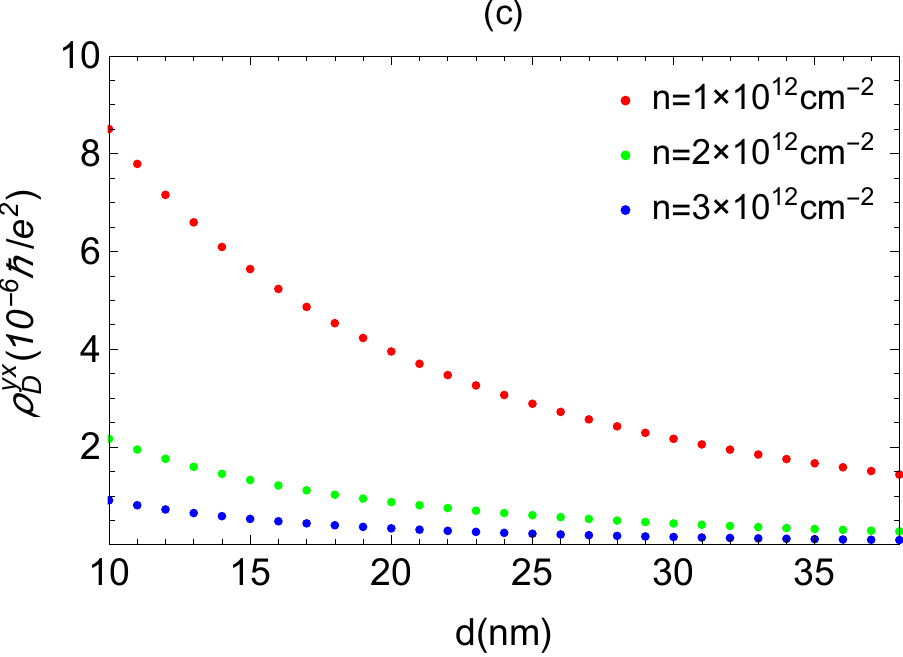}
\caption{\label{yx3} (a) Electron density dependence at $T=5~\mathrm{K}$, (b) Temperature dependence with $n=10^{12}\text{cm}^{-2}$, (c) Layer separation dependence at $T=5~\mathrm{K}$. $M_{\text{p}} =10~\mathrm{meV}$, dielectric constant $\epsilon_r=20$, $A=4.1~\text{eV\AA}$ and transport time $\tau\approx0.1~\mathrm{ps}$.}
\end{center}
\end{figure}

Experimentally, for TI films in the large-surface limit, non-topological contributions from the bulk and the side surface are negligible, \cite{TME_TI_PRB_Nagaosa} and we expect the effects described in this work to be observable. Aside from commonly used materials such as Bi$_2$Se$_3$ and Bi$_2$Te$_3$, a small-gap three-dimensional TI has also been identified in strained HgTe. The TI surface states in this material are, however, spatially extended and could be peaked quite far from the wide quantum well edges, reducing the effective 2D layer separation. \cite{Tilahun_TIF_QHS_PRL2011}

Our theory predicts that in a sample in which the active layer is undoped, so that it produces no longitudinal current but only a quantized anomalous Hall current, there will be no drag current at all in the passive layer. Whereas this is in agreement with Ref.~[\cite{Drag_1D_Fiete}] its subtleties warrant a separate discussion. In this context, we note that the description of quantum Hall effects can be formulated in terms of a bulk model that accounts for the Berry phase, as well as in terms of an edge state model. The two must evidently agree. In the latter context Ref.~[\cite{Drag_1D_Fiete}] demonstrated that for chiral edge states in quantum spin-Hall systems this charge imbalance does not translate into a drag current due to the absence of backscattering. Quantum anomalous Hall systems, which have a single edge state, are a subset of quantum spin-Hall systems, which have two chiral edge states. The QAHE can also be understood as a charge imbalance between chiral edge states. Hence one expects the drag current to vanish quite generally at zero doping in quantum anomalous Hall systems. The model used in this work, which is also applicable to quantum spin-Hall systems, can be used to calculate explicitly the edge contribution (Supplement). In complete agreement with the bulk formulation described above, when our model is applied to the edge states it yields zero drag. Yet the edge states are poorly defined at high doping, $\varepsilon_\text{F} \tau_l/\hbar \gg 1$, which is the focus of this work. In addition, the Dirac fermions studied here live in TIs, which have only one shared edge between the top and bottom surfaces, making it tricky to describe thicker samples using an edge state model. In fact, at zero doping, the TI drag experiment is not well defined, since in that parameter regime leakage into the sidewall states is unavoidable.

Experimentally, for TI films in the large-surface limit, non-topological contributions from the bulk and the side surface are negligible \cite{TME_TI_PRB_Nagaosa}, and we expect the effects described in this work to be observable. Aside from commonly used materials such as Bi$_2$Se$_3$ and Bi$_2$Te$_3$, a small-gap three-dimensional TI has also been identified in strained HgTe. The TI surface states in this material are, however, spatially extended and could be peaked quite far from the wide quantum well edges, reducing the effective 2D layer separation \cite{Tilahun_TIF_QHS_PRL2011}.    

It is well known that the undoped limit of a magnetic topological insulator is a special case. \cite{Culcer_TI_AHE_PRB11} When the chemical potential lies in the middle of the gap opened by the magnetization between the surface conduction and valence bands the anomalous Hall conductivity is quantized: in three dimensions a single TI surface contributes exactly $e^2/(2h)$ to the Hall conductivity, referred to as the quantized anomalous Hall effect (QAHE). In a doped system the QAHE effectively yields an offset to the measured anomalous Hall conductivity. In the picture presented here the QAHE emerges naturally as the special case in which the carrier density in the conduction band is taken to zero. The effective two-dimensional model we have used throughout this work predicts that in the special case of the quantum anomalous Hall effect, when the system is undoped, the surface conduction band is empty, and the chemical potential lies in the magnetization gap, the drag current is identically zero. In the previous section we have discussed above the physical interpretation of this result that emerges from our effective two-dimensional picture. It is known, however, that the quantum anomalous Hall effect is associated with a set of chiral edge modes which are well-defined at zero doping (and only then). In this subsection we demonstrate that an effective one-dimensional model for the edge modes at zero doping yields the same result as the two-dimensional model, and gives additional physical insight. 

To this end we consider the broader case of Coulomb drag between two identical quantum spin-Hall systems, each with one Kramers pair on its edge. A current $I_1$ is driven along the upper edge of the lower quantum spin-Hall system and through electron-electron interactions a voltage $V_2$ is induced in the lower edge of the upper quantum spin-Hall system. Each quantum spin-Hall edge state can be described by Hamiltonian $H_{k} =Ak_x\sigma_z$, and the edge dispersion is $\epsilon_k^{\pm}=\pm A k_x$. Hence impurity scattering term $\hat{J}_0(f^{(l)}_{\bm k})$ becomes
  \begin{equation}
 \ba 
 P_{\parallel}\hat{J}(f^{(l)}_{{\bm k}\parallel})&\dps =\frac{n_i}{\pi\hbar A}\int d{k'_x}|\overline{U}_{{\bm k}{\bm k}'}|^2 (f^{(l)}_{{\bm k}\parallel}-f^{(l)}_{{\bm k}'\parallel})\sigma_z\delta(k_x-k'_x),
\\[3ex]
\ea
 \end{equation}
where ${\bm k}=k_x\hat{\bm e}_x$  for 1D-case and $P_{\parallel}\hat{J}(f^{(l)}_{{\bm k}\parallel})=0$. In the meantime, the direct inter-layer electron-electron scattering term of Eq.~(\ref{electron-electron-m}) becomes

 \begin{equation}\label{e-e-1D}
 \ba
  J^i(f_{\bm k})_{s_{\bm k}} \dps= & \dps- \frac{2\pi}{\hbar}\!\sum_{{\bm k}'}\!|v_{|{\bm k}-{\bm k}_1|}^{(\text{pa})}|^2F^{\text{(p)}}_{s_{\bm k}s_{{\bm k}_1}} F^{\text{(a)}}_{s_{{\bm k}'}s_{{\bm k}'_1}} \\ [3ex]
  
  &\dps \delta[\varepsilon^{(\text{p})}_{k_1,s_{{\bm k}_1}}\!\!-\!\varepsilon^{(\text{p})}_{k,s_{\bm k}}\!+\!\varepsilon^{(\text{a})}_{k'_1,s_{{\bm k}'_1}}\!-\!\varepsilon^{(\text{a})}_{k',s_{\bm k'}}\!] \\ [3ex]
  
  &\dps \times \big\{f^{(\text{p})}_{{\bm k},s_{\bm k}}[1\!-\!f^{(\text{p})}_{{\bm k}_1,s_{{\bm k}_1}}]f^{(\text{a})}_{{\bm k}',s_{\bm k'}}[1\!-\!f^{(\text{a})}_{{\bm k}'_1,s_{{\bm k}'_1}}] \\ [3ex]
  
  & \dps - [1\!-\!f^{(\text{p})}_{{\bm k},s_{\bm k}}]f^{(\text{p})}_{{\bm k}_1,s_{{\bm k}_1}}[1\!-\!f^{(\text{a})}_{{\bm k}',s_{\bm k'}}]f^{(\text{a})}_{{\bm k}'_1,s_{{\bm k}'_1}}\!\big\}.
\ea
\end{equation}

If the distribution function is the equilibrium one, $J^i(f_{\bm k}) _{s_{\bm k}}=0$. In QSH 1D-case, the wave function overlap $F^{(l)}_{s_{\bm k}s_{{\bm k}_1}}=\langle s_{\bm k}l|s_{{\bm k}_1}l\rangle\langle s_{{\bm k}_1}l|s_{\bm k}l\rangle=1$ with $s_{\bm k}=s_{{\bm k}_1}$ and $F^{(l)}_{s_{\bm k}s_{{\bm k}_1}}=\langle s_{\bm k}l|s_{{\bm k}_1}l\rangle\langle s_{{\bm k}_1}l|s_{\bm k}l\rangle=0$ with $s_{\bm k}\neq s_{{\bm k}_1}$. We found $ J^i(f_{\bm k})_{s_{\bm k}}=0$. Because the spectrum is linear there is no contribution to the drag from forward scattering, and backscattering is forbidden by time-reversal symmetry. \cite{Drag_1D_Fiete} Hence the Coulomb drag between two identical chiral quantum spin-Hall  systems is identically zero. \cite{Drag_1D_Fiete} 

The argument that chiral edge states do not give rise to drag is general and applies just as well to the Hall conductivity as to the longitudinal conductivity. The key physics is the absence of backscattering. In chiral quantum spin-Hall systems the drag current is zero because backscattering is forbidden by the linear Dirac-like dispersion. In this sense quantum anomalous-Hall systems can be regarded as a special case of quantum spin-Hall systems: they have only one state on each edge, rather than a Kramers pair. So even if backscattering were allowed by the quasiparticle dispersion there would be no states to backscatter into. Hence the drag current in quantum anomalous-Hall systems is identically zero. This result applies also to helical Fermi liquids, such as HgTe quantum wells and thin Bi$_2$Se$_3$ in the 2D limit. Yet the edge states are poorly defined at high doping, $\varepsilon_\text{F} \tau_l/\hbar \gg 1$, which is the focus of this work. In addition, the Dirac fermions studied here live in TIs, which have only one shared edge between the top and bottom surfaces, making it tricky to describe thicker samples using an edge state model. In fact, at zero doping, the TI drag experiment is not well defined, since in that parameter regime leakage into the sidewall states is unavoidable. 

\section{Beyond topological insulator films}
\label{sec:beyond}

The focus of the paper up to now has been on thin films of topological band insulators in which both layers are doped with the same type of carrier, that is, either electron-electron or hole-hole layers. Our conclusions can straightforwardly be extended to structures beyond those considered thus far. We will now discuss the necessary modifications for treating the cases of multi-valley TIs and ultra-thin films, the possibility of exciton condensation, and massive Dirac fermions beyond magnetic topological insulators.

Certain materials, such as the topological Kondo insulator SmB$_6$, have more than one valley. In this case, a valley degeneracy factor $g_v$ will need to be introduced.
In the following, we will fix the electron density $n$. The polarisation $\Pi \propto \sqrt{g_{v}},$ while the susceptibility $\chi \propto g_{v}$, thus the screening function
$\epsilon(q,\omega)$ is proportional to $g_{v}$ \cite{SDS_Gfn_RMP11}. The drag current, and hence the drag conductivity, will remain the same as in the single-valley case. For the drag resistivity, based on Eq. (\ref{ll-result}) we will get the linear $g_{v}$ dependence because $\sigma_{a,p} \propto 1/\sqrt{g_{v}}.$ Note that there will be an intervalley
impurity scattering contribution if short-range disorder is present in the system, but this will simply result in a renormalisation of the scattering time tau by an intervalley scattering term.

The case of ultra-thin films deserves special attention. When the TI film is ultra thin tunnelling is enabled between the top and bottom surfaces. The tunnelling between the surface states on the top and bottom surfaces may open an energy gap in the energy spectrum \cite{Weizhe_TITF_2014_prb}. In this case, the massless Dirac Hamiltonian $H_{0{\bm k}}^{(l)}$ needs to be augmented by a series of tunnelling terms, and is generally written as:
\begin{equation}
    H_{\bm k}=A\tau_z\otimes[{\bm\sigma}\cdot({\bm k}\times\hat{\bm z})] + t\tau_x\otimes \mathbb{1},
\end{equation}
where $\tau$ matrices represent the layer pseudospin space, with $\tau_z=1$ symbolising the up surface and $\tau_z=-1$ the bottom surface. Here $\hat{\bm z}$ is the unit vector in the direction of $\bm z$, and the term $t$ represents the tunnelling matrix element between two opposite topological surfaces. After the direct diagonalization, the energy spectrum of the TI thin film is given by $\epsilon_{\bm k}=\pm\sqrt{t^2+A^2k^2}$, which has a gap of size $2t$. The disorder scattering term becomes:
\begin{equation}\label{Jt}
    \ba
    \dps\hat{J}(f_{\bm k})=&\dps\frac{n_i\epsilon_k}{4\pi A^2\hbar}\int_0^{2\pi} \, d\theta_{\bm k'} \, |U_{{\bm k}{\bm k'}}|^2(f_{\bm k} - f_{\bm k'}) \\[3ex]
    &\dps\times\bigg(1 + \frac{t^2}{t^2+A^2k^2} + \frac{A^2k^2\cos\gamma}{t^2+A^2k^2}\bigg),
    \ea
\end{equation}
where $\gamma = \theta_{\bm k} - \theta_{\bm k'}$ is the angle between the incident and scattered wave vectors. Note that the density matrix entering this term is the full density matrix $f_{\bm k}$ of the double-layer system, rather than its projection onto each individual layer. Two limiting cases can be identified: weak tunnelling $t \ll \varepsilon_\text{F}$ and strong tunnelling $t \gg \varepsilon_\text{F}$.

For weak tunnelling, $t \ll Ak_F$, the carrier wave functions are overwhelmingly located in one of the two layers (surfaces), and the notion of Coulomb drag can be retained to a good approximation. The effect of tunnelling can be taken into account perturbatively. For example, the momentum relaxation time becomes $(1/\tau) \rightarrow (1/\tau) \, [ 1 + t^2/(A^2k_{\text{F}^2}) ]$. Interlayer tunnelling slightly decreases the momentum relaxation time. Screening is qualitatively different \cite{Weizhe_TITF_2014_prb}, but for weak $t$ it is a good approximation to retain the screening function defined in Eq.\ (\ref{dielectric}). We expect, rigorously, $\rho_{\text{D}} \rightarrow \rho_{\text{D}} [ 1 - 2t^2/(A k_F)^2 ]$. The numerics display a similar trend.
 
For strong interlayer tunnelling $t \sim A k_F$ the notion of Coulomb drag is not applicable to the thin film system. In this case, the carrier wave functions spread over the two layers, hence the picture of the Coulomb interaction causing charges in one layer to \textit{drag} charges in the other is no longer valid. The main effect of electron-electron interactions will be through the Coulomb renormalization of the conductivity \cite{Weizhe_TITF_2014_prb}. We note, however, that the film needs to be extremely thin for the tunnelling gap to be noticeable \cite{LuShan_TITF_MassiveDirac_spinphys_PRB2010}, therefore we expect realistic samples to lie in the weak tunnelling limit.

When the active layer is doped with electrons and the passive layer is doped with holes, or vice-versa, exciton condensation may occur. This effect is driven by an exchange term in the interlayer Coulomb interaction. In principle, for interlayer exchange to be nonzero tunnelling also has to be nonzero, but one can think of a situation in which the tunnelling is negligible but the exchange is not. The interesting problem concerning the way exciton condensation impacts $\rho_{xx}$ has been considered in great detail in Ref.~\cite{Polini_Exciton_Drag_PRL2012}, where it was shown that the drag resistivity exhibits an upturn at low temperatures described by a logarithmic dependence on the temperature.

The theory of Ref.~\cite{Polini_Exciton_Drag_PRL2012} has recently been shown to be a good description of experimental observations in graphene \cite{Pellegrini_Exciton_NCom2014}, where an exciton condensate phase has been identified, with the critical temperature estimated at 10 - 100mK. This estimate was for a sample grown on a GaAs substrate, in which the effective dielectric constant is expected to be $\epsilon_r \approx 6$, whereas in current TI films the lowest experimentally reported $\epsilon_r \approx 30$ \cite{Beidenkopf_NP11}, largely due to screening by the unavoidable bulk of the film. These observations suggest that the critical temperature in TI films could be at least an order of magnitude smaller than in graphene, placing it in the range 1 - 10 mK, which would make exciton condensation rather difficult to detect experimentally in currently available samples.

Massive Dirac fermions are also found in graphene with a staggered sublattice potential, and $\text{MoS}_2$ thin films where inversion symmetry is broken. The band Hamiltonian for a single layer is given by $H^{(l)}_{0{\bm k}}=at(\tau k_x\sigma_x+k_y\sigma_y)+\frac{\Delta}{2}\sigma_z$ with $\tau=\pm1$ the valley index, where $a$ is the lattice constant, $t$ is the effective hopping integral, and $\Delta$ is the energy gap. These Dresselhaus-like Hamiltonians can be directly mapped onto the Rashba Hamiltonian considered in this work \cite{DM_Balatsky_2014}. The longitudinal drag current is identical for both TIs and other massive Dirac fermion systems, because the physical mechanism behind the longitudinal drag phenomenon is a result of rectification by the passive layer of the fluctuating electric field generated  by the active layer. However, when inversion symmetry is broken in a 2D hexagonal lattice, a pair of valleys which are time-reversal of each other are distinguishable by their opposite values of magnetic moment and Berry curvature. Therefore, there will be no transverse drag current in graphene or $\text{MoS}_2$ with broken inversion symmetry because the Berry curvatures have opposite signs in the opposite valleys. In a Dirac semimetal, each Dirac point is four-fold degenerate and can be viewed as consisting of two Weyl nodes with opposite chiralities. Consequently, transverse drag currents also vanish in Dirac semimetals. At small magnetizations the longitudinal drag currents in these materials will be independent of the magnetizations of either layer. 
    
The Hall drag results discussed in this work are also applicable to Rashba 2DEGs, though measuring a strong anomalous Hall effect \cite{Culcer_AHE_PRB03} can be challenging. A sizable fraction of the conductivity quantum is obtained if the two Rashba sub-bands experience a large magnetization splitting and $\varepsilon_\text{F}$ lies in the bottom sub-band, but that is challenging experimentally. When $\varepsilon_\text{F} \gg M$ the effect vanishes altogether. 
    
\section{Conclusions}
\label{sec:con}

We have discussed a complete picture of Coulomb drag in Dirac fermion systems incorporating both massless and massive Dirac fermions. For massless Dirac fermions the longitudinal drag resistivity is proportional to $T^2d^{-4}n^{-3/2}_{\text{a}}n^{-3/2}_{\text{p}}$ at low temperature and electron densities. We have shown that the drag effect is expected to be weaker in topological insulators than in graphene, and that different regimes are accessible experimentally in these two massless Dirac fermion systems. For massive Dirac fermions we have shown that the drag due to the topological terms on the active surface vanishes and the only contribution to anomalous Hall drag comes from the anomalous Hall current generated by the longitudinal drag force experienced by the passive layer. In a system in which the active layer is undoped and only a quantized anomalous Hall effect exists in the active layer there is no drag current at all. In a doped system the transverse drag current has a non-monotonic dependence on the magnetization of the passive layer, with a peak at a value of the magnetization that becomes pronounced at lower densities.

\section{acknowledgments}
This research was supported by the Australian Research Council Centre of Excellence in Future Low-Energy Electronics Technologies (project number CE170100039) and funded by the Australian Government. The authors thank Di Xiao, Haizhou Lu, Yaroslav Tserkovnyak, Marco Polini, Greg Fiete, Vladimir Zyuzin, A. H. MacDonald, Oleg Sushkov, Tommy Li, Euyheon Hwang, Zhenyu Zhang, Wenguang Zhu, Zhenhua Qiao, Changgan Zeng, Shun-Qing Shen, and W.~K. Tse for enlightening discussions.

\section*{References}
 
\end{document}